# Transforming ASN.1 Specifications into CafeOBJ to assist with Property Checking

| Konstantinos Barlas | George Koletsos | Petros Stefaneas |
|---|---|---|
| School of Electrical and Computer Engineering | School of Electrical and Computer Engineering | School of Applied Mathematical and Physical Sciences |

National Technical University of Athens
Polytexneioupoli Zografou, Iroon Politexneiou 9, 15780, Athens, Greece

| +306977359857 | +302107721773 | +302107721869 |
|---|---|---|
| kosbarl@central.ntua.gr | koletsos@math.ntua.gr | petros@math.ntua.gr |

## ABSTRACT
The adoption of algebraic specification/formal method techniques by the networks' research community is happening slowly but steadily. We work towards a software environment that can translate a protocol's specification, from Abstract Syntax Notation One (ASN.1 – a very popular specification language with many applications), into the powerful algebraic specification language CafeOBJ. The resulting code can be used to check, validate and falsify critical properties of systems, at the pre-coding stage of development. In this paper, we introduce some key elements of ASN.1 and CafeOBJ and sketch some first steps towards the implementation of such a tool including a case study.

## Categories and Subject Descriptors
D.2.4 [**Software Engineering**]: Software/Program Verification – *correctness proofs, formal methods, validation.*

## General Terms
Design, Reliability, Security, Standardization, Languages, Verification.

## Keywords
Specifications, Abstract Syntax Notation One, ASN.1, CafeOBJ, formal verification, algebraic specification, correct translation, protocol specification, property checking.

## 1. INTRODUCTION
Software reliability is a critical aspect of software quality. A way to ensure reliability is to make full use of formal specifications. Using formal methodologies we can make sure that the resulting code is a correct implementation of the specification. Such methods can be applied, for example, at the stage before a program is being tested in real time, maximizing the chances that it will be error-free when it gets released. Telecommunication protocol standards have been written in the Abstract Syntax Notation One (ASN.1 – a data definition language) format, however there's little we can do about the verification of properties in this format. We have been working towards a software environment that can translate a protocol's specification that is written in the ASN.1 format, into a fully executable set of CafeOBJ [Diaconescu and Futatsugi 1998] specifications. The resulting code can then be used to verify critical properties of the designed system via the powerful CafeOBJ proving engine. There have been some related works, such as the work of Senachak, et al. [2005] which combines the specification language CafeOBJ with Java. Another similar approach can be found in the work of Martin and Thomas [1994]. The authors have investigated the verification requirements of LOTOS specifications and determined the applicability of equational reasoning and term rewriting to discharging these requirements. They also developed a tool for ASN.1 to LOTOS translation.

Our approach is different from the above work, since it aims more towards system verification and/or falsification, hoping that it can be used as a great asset to protocol developers; we can take advantage of code already written in ASN.1 and by properly translating it into CafeOBJ, we can use the proof engine for property checking.

A approach to ours can be found in the work of Shanbhag et al. [2001]. It integrates ASN.1 and model checking by creating an enhanced version of ASN.1 (EASN). EASN is created by replacing SPIN's model checking tool datatyping abilities with ASN.1 data types. This paper is an improvement of previous versions, as seen in the works of Barlas et al. [2009] and Barlas et al. [2010].

## 2. ASN.1

### 2.1. Introduction to ASN.1
ASN.1 is a joint ISO/IEC and ITU-T standard and is a well-known notation used in describing messages to be exchanged between communicating application programs [Dubuisson 2001; Larmouth 2001]. It is a framework for representing tree structured data, providing a set of formal rules for describing the structure of objects that are independent of machine-specific encoding techniques (freeing protocol designers from having to focus on the bits and bytes layout of messages) and is a precise, formal notation that removes ambiguities.

An ASN.1 specification does not specify by itself the concrete representation of the data in transmission or storage. The concrete representation is stated as the transfer syntax in ASN.1 terms and is determined by the application of one of a set of encoding rules.

An obvious advantage of ASN.1 is that the same data (semantically) can be formatted in other various other ways (syntactically) by applying different encoding rules. For example, a very compact binary representation uses PER [Dubuisson 2001]. Currently the only "transfer syntax" for XML Schema or RELAX NG is XML.

ASN.1 has been initially used to describe email messages within the Open Systems Interconnection protocols, and since then has been adopted for use by a wide range of other applications, such as in network management, secure email, cellular telephony, air traffic control, voice and video over the Internet, security, authentication & cryptography, banking, wireless networks, transportation, energy, electronic tags and cards, health and genetics and graphics and file transfer.

## 2.2. ASN.1 Extensions

ASN.1 is designed for efficiency and the data is usually packed into byte boundaries, and hence is not very readable and is not very easy to manipulate. Since ASN.1 data is structured data, it is possible to represent the same information in Extensible Markup Language (XML). XML is not particularly efficient in terms of data length, but is more readable, and it has many off-the-shelf free tools (e.g., XML processors for parsing and generation, XSL processors for rendering, XML editors for authoring, and so on). XML can serve as transfer syntax for ASN.1, and the cooperation of those two notations can only benefit both of them.

What makes ASN.1 so important is that there are numerous examples of existing, widely deployed, mature applications defined in terms of ASN.1, and a whole new XML-keen group that is more than happy to use XML (or XML schema language) as means of formal specification.

IBM (among other companies) has developed a two way ASN.1 to XML translator (ASN.1/XML translator) as part of its XML Security Suite, providing a Java library for the translation. Also the ASN.1 Project from ITU-T started working (jointly with ISO) on two XML-related initiatives.

Also, ASN.1 conversion to C has been made possible, and lots of different tools exist on the market (ASN.1 to C/C++ Compiler, ASN.1 to Java/C# Compiler) that generate a full set of a working C program generated from a set of ASN.1 specification modules. Also, there's a cross compiler from ASN.1 to Erlang (Asn1ct) which generated encode and decode functions to be used be Erlang programs sending and receiving ASN.1 specified data. There is also a set of ASN.1 tools for Python, (ASN.1 tools for python), which implements ASN.1 data types (concrete syntax) and codecs (transfer syntaxes) for Python programming environment.

## 2.3. Basic Syntax of ASN.1

Here we will display some basic ASN.1 syntax rules that will allow the reader to understand the case study that follows. Below is a small excerpt of a larger specification module that describes a banking account. The module "Account" will be used to define the data exchanged for the central entity of a banking account system –the bank account -. We can spot how the ASN.1 keywords are always spelt in capital letters except some character string types such as NumericString).

```
Account ::= SEQUENCE {
 iban    NumericString (SIZE (27)),
 client  Client,
 balance Balance  }
```
**Figure 1. Example of SEQUENCE**

We can read the above as: 'an Account is a Sequence of three types: the first called "iban" is used to denote numeric data of a fixed length (27 digits); the second component is called "client" and denotes data of type "Client" and so on.

"Account" is an ASN.1 type; notice that it begins with a capital letter, it is followed by the "::=" symbol and that its definition does not end with a semicolon contrary to many programming languages.

"SEQUENCE" is the most basic and mostly used ASN.1 data type, denoting nested data. Alternatives to SEQUENCE are SET, SET OF and SEQUENCE OF. Semantically all four types represent the same thing, structured data. One similar and equally common found data type in ASN.1 specifications is 'CHOICE'. Choice only allows one of the nested components to be transmitted, and it's very useful when a choice has to be made. Its syntax is similar to the Sequence data type.

```
Payment-method ::= CHOICE {
 check       Check-number,
 credit-card SEQUENCE { number Card-number,expiry-
date Date}}
```
**Figure 2. Example of CHOICE**

The above examples can be read: 'The Payment-method is a choice (CHOICE) between these two components: the first called check, denoting data of type Check-number (which of course is similarly denoted later); the second called credit-card, denotes a sequence of two more components; a component named number of type Card-number, and a component called expiry-date of type Date.

Please note that ASN.1 has no knowledge of the data types beyond the ones built-in, so we have to write further details for the types 'Client' and 'Balance'.

```
Client ::= SEQUENCE {
 clientid Integer,
 firstname PrintableString (SIZE (1..30)),
 lastname PrintableString (SIZE (1..50)),
 street PrintableString (SIZE (1..50)),
 postcode NumericString (SIZE (5)),
 city PrintableString (SIZE (1..30)),
 country PrintableString (SIZE (1..20))  }
```
**Figure 3. A client's data**

This can be read: 'A Client, is a structure of seven (we can of course add more in the same way) components; The clientid (an integer), the first and last name, with a maximum length of 30 and 50 characters accordingly and finally (denoted in the same manner) street, postcode, city and country'. The Balance module is declared accordingly.

Last but not least, ASN.1, as a specification language, prescribes no order when defining the entities. It is actually possible to reference the data types before their definition in the specification. It is a good idea for readability issues to apply a top-down approach to our problem; from general to particular. [Dubuisson 2000]

## 3. CAFEOBJ:

## 3.1 An Algebraic Specification Language

CafeOBJ is a new generation algebraic executable, industrial strength algebraic specification language/system. The main underlying logics of CafeOBJ are order-sorted algebras

[Diaconescu et al. 1998; Goguen et al. 1992] and hidden algebras [Diaconescu et al. 1998; Diaconescu et al. 2000]. The former is used to specify abstract data types while the latter to specify abstract state machines, providing support for object oriented specifications.

It is a specification and programming language. Since it's a direct successor of OBJ, it inherits all its features (flexible mix-fix syntax, powerful typing system with sub-types, and sophisticated module composition system featuring various kinds of imports, parametrised modules, views for instantiating the parameters, module expressions, etc.) but it also adds rewriting logic and hidden algebra, as well as their combination.

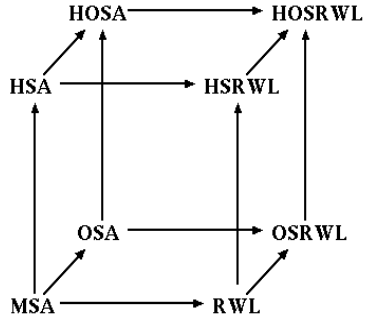

**Figure 4. CafeOBJ Cube**

Figure 4 displays how CafeOBJ uses institution embeddings (represented by arrows) to combine several logics such as many sorted algebra, order sorted algebra, hidden algebra and rewriting logic. (M for many, O for ordered, H for hidden, S for sorted, A for algebra and RWL for rewriting logic.)

Each CafeOBJ module consists of three main parts: sort declaration, operation declaration and axiom declaration.

There are two kinds of sorts in CafeOBJ, *visible sorts* representing abstract data types, and *hidden sorts* representing the set of states of a state machine. The operations to hidden sorts are classified into *actions, observations* and *hidden constants*. An action can change the state of an abstract machine. It takes a state of the abstract machine and zero or more data, and returns another or the same state of the abstract machine. An observation is used to observe the value of a data component in an abstract machine. It takes a state of an object and zero or more data, and returns the value of a data component in the abstract machine. Finally, the hidden constants denote initial states of abstract machines.

## 3.2 Syntax of CafeOBJ
CafeOBJ notation is presented through the example of a list of natural numbers:

```
mod! NATLIST {
  pr(NAT)
  [NatList]
  op nil : -> NatList
  op _|_ : Nat NatList -> NatList
  --
  op _@_ : NatList NatList -> NatList
  op mkl : Nat Nat -> NatList
  --
  vars X Y : Nat
  vars L1 L2 : NatList
  --
  eq nil @ L2 = L2 .
  eq (X | L1) @ L2 = X | (L1 @ L2) .
  --
  eq mkl(X,Y)
    = if Y < X then nil else X | mkl(X + 1,Y) fi.}
```

**Figure 5. A sample CafeOBJ module**

The keyword mod! indicates that the module is a tight semantics declaration. Visible sorts are declared by enclosing [ and ]. NatList is the visible sort of the lists of natural numbers. pr(NAT) denotes protecting import of module NAT, that is a built module specifying natural numbers. Apart from protecting imports, which do not collapse elements or add new elements to the models of the imported modules, there exist extending imports, denoted by ex, that may add new elements but not collapse elements, and using imports, denoted by using keyword, that provide no guaranty, so they might even collapse elements. The keyword op or ops is used to declare non hidden operators. The operator nil is a constant denoting the empty list. The operator _|_ takes a natural number and a list and returns the new list, while the operator _@_ denotes the concatenation of two lists. Finally, the function mkl(m,n) makes the list m | m+1| … | n | nil for two natural numbers m, n. If n < m then it returns the empty list. The lines beginning with the keyword -- are comments. Keyword var (vars) declares CafeOBJ variables, while the equations defining axioms of the specification begin with eq or ceq when the equation is conditional. First and second equations defines the concatenation of an empty or non-empty list with a non-empty list, respectively. The final equation defines function mkl. The CafeOBJ system uses declared equations as left-to-right rewrite rules and reduces a given term. This makes it possible to simulate specified systems and verify that they have desired properties. In the case of NATLIST, some simulations that we can perform are as follows:

```
open NATLIST
  red nil @ nil .
  red nil @ (0 | 1 | 2 | nil) .
  red (0 | 1 | 2 | nil) @ (3 | 4 | 5 | nil) .
  red mkl(0,10) .
  red mkl(5,5) .
  red mkl(5,4) .
close
```

**Figure 6. Red command in CafeOBJ**

The command open makes a temporary module that imports a given module and close destroys it. The command red reduces a given term.

## 3.3 OTS
Observational Transition System [Ogata et al. 2006] is a transition model that defines the behaviour of a system as *actions* happen (changes on observation values after applying a sequence of transitions at any state of system). Writing a specification in OTS format gives the advantage of specifying a system in an abstract way.

Assuming that 'Y' is the universal data space (declared by a hidden sort) and '$D_*$' the data types used in OTSs [Ogata et al. 2006], an OTS 'S' is a set of {O,I,T} such that: i) 'O' is a finite set of observers from Y to $D_*$ (each operator $o \epsilon O$ is a function from Y to D) ii) 'I' is the set of initial states such that I is a subset of Y and iii) 'T' is a finite set of transitions $\tau$ from Y to Y (transitions are indexed functions that 'move' the system from one state into another). Each transition $\tau$ creates a *successor state* of Y with respect to $\tau$. The condition $c_\tau$ that is required for a transition rule $\tau \epsilon T$ (which is predicate on states) to take place is called the

*effective condition*. [Senachak et al. 2005]

# 4. TRANSLATING ASN.1 SPECIFICATIONS INTO CAFEOBJ

Over the past years ASN.1 enjoyed the support of a very important consortium and it will continue to be one of the top specification languages. On the other hand CafeOBJ, is relatively new, slowly emerging as a specification language and mostly as means of property checking.

Our idea aims at helping a designer, who wants to take advantage of an already written ASN.1 specification of the designed system, in order to check if the system holds some critical properties. Without that, the process would be to rewrite a specification from scratch into a language that supports proofing methodologies, something that could cost lots of time.

Our suggestion consists of a software environment that takes an existing ASN.1 specification and transforms it into a set of CafeOBJ modules, giving as a result a foundation we want (a basic set of modules representing data) which can be used to further test the protocol, checking the validity of critical system properties. So in this way, a designer can automatically come up with basic CafeOBJ modules, and should only have to deal with system checking instead of rewriting and checking.

## 4.1. Differences Between the Two Environments

ASN.1 and CafeOBJ are both used for system specifications, but in a very different way as they can only be used to specify different aspects of the system; ASN.1's primary use is to specify the static structure of the data that is handled by a system and the signatures of the system's provided operations. But, it can't be used to specify any part of a system's dynamic behavior since it has no means of understanding system states.

So, we can't use ASN.1 to specify how an operation may change system's data or the actual changes that may occur to data.

On the other hand, CafeOBJ, can handle both of those cases easily, by using 'actions' that specify the type of changes that can happen on a system, and with the use of 'observations' that return data after an action has taken place. CafeOBJ has a different way to specify both of these, as its high level of abstraction means that the structure of the data is actually implied by the specification (in terms of abstract sorts) instead of being explicitly described as in ASN.1. This means that any translation from ASN.1 to CafeOBJ cannot be 100% complete; the output will only be a part of the equivalent CafeOBJ specification, still lots of hard work is already done automatically, and now all we would have to do is to manually provide some additional information and code rewriting to further proceed with the system property checking.

## 4.2 The Need for Verified Transformation Rules

To accomplish our goal of transforming the numerous ASN.1 specifications into CafeOBJ modules, we will need to set some ground rules. Rules that at any given time and for any given ASN.1 module, can translate the structure from one language to another while making sure that no critical data could be either lost or misinterpreted. Therefore those transformation rules will have to be logically proved, so we can ensure the correctness of the engine.

**Rule 1:** An ASN.1 module name serves as a file in CafeOBJ, containing all of the modules.

```
ASN.1: mName {arguments} DEFINITIONS
{arguments} :=
            BEGIN
                …
            END
CafeOBJ: mName.mod
```

**Rule 2:** Each ASN.1 type name serves as a module in CafeOBJ.
```
ASN.1: Typename ::= SEQUENCE {
                    ...}
CafeOBJ: module* TypeName { }
```

**Rule 3a:** ASN.1's most used data structures are 'sequences'. Each Sequence name (which acts as a place holder for the nested data types) is being modeled as a CafeOBJ sort and all of the nested types become sub sorts of the main Sort. This way we preserve the semantics of the original declarations.
```
ASN.1:
Typename ::= SEQUENCE {
  subtypename₁    sometype),
  subtypename₂    sometype,
...,
  subtypenameₙ    sometype }
CafeOBJ:
module* TypeName {
[ Typename > subtypename₁ subtypename₂ …
subtypenameₙ ] }
```
This looks a lot like something that could benefit from the use of CafeOBJ's record type. However, the way records are declared in CafeOBJ allow fields (slots) to be optional and we only want this to happen when an ASN.1 type has the "OPTIONAL" mark.

Our approach on the 'choice' data structure, would be similar to "Sequence " translation but on top that we add a few operators that only allow one of the viable options to be actually transmitted. Something along the lines that only one of the available parameters to choose from can actually be non-null, otherwise CafeOBJ will not classify the returning state as a reachable one. Data types such as SET, SET OF and so on are modelled in the exact same way.

**Rule 4:** To import an ASN.1 type from a **different** ASN.1 module then in our CafeOBJ file we will declare naturally that type as a CafeOBJ module and then every CafeOBJ module in our current file (which corresponds to every ASN.1 type in our current module) will import that CafeOBJ module. There is also an "EXPORTS" clause in ASN.1, dictating us which ASN.1 types of the current module can become available to other modules. Omission of an "EXPORTS" statement in ASN.1 means that "everything is available for import by another module" whilst "EXPORTS ;" has the semantics "nothing is available for import by another module". [Larmouth 2001]. We can safely ignore any sort of "EXPORTS" clause while translating into CafeOBJ simply because it is being implied by a correct use of "IMPORTS" clause.

```
ASN.1 (while inside mName):
BEGIN
IMPORTS type₁₋₁, type₁₋₂, …, type₁₋ₙ FROM mName₁
Type₂₋₁, type₂₋₂, …, type₂₋ₘ FROM mName₂ ;
CafeOBJ (in mName.mod we):
```
- Declare the modules type₁₋₁, …, type₂₋ₘ
- Every module in mName.mod will protect those modules

```
Protecting (type₁₋₁ + type₁₋₂ + … + type₂₋ₘ)
```

**Rule 3b:** For sequences inside sequences we can now extend Rule 3a using Rule 4. We can specify the nested sequence first and import the resulting module into the bigger one.

This is everything that we can derive directly from an ASN.1 specification. However to be able to work easily with types and subtypes in CafeOBJ we will need a few more work. When we reach the verification process we have to be able to access every piece of information that the original ASN.1 specification provides structure for. To assist us with verification purposes, we will add some extra operators (observational operators) in each CafeOBJ module:

**Rule 5a** (Operator declaration):
We create one operator with the module's name that gets as input all of the sub sorts and returns the main sort type. Also, for each of the sub sorts we create an operator that given the main sort as input, it returns the specific sort type. These operators are named *returnXXX*, where XXX is the sub sort name.

**ASN.1:**
```
Typename ::= SEQUENCE {
 subtypename₁  sometype),
 subtypename₂  sometype,
 ...,
 subtypenameₙ  sometype }
```
**CafeOBJ:**
```
op typename : subtypename₁ subtypename₂ ...
subtypenameₙ -> Typename
op returntypename₁ : Typename -> subtypename₁
op returntypename₂ : Typename -> subtypename₂
...
op returntypenameₙ : Typename -> subtypenameₙ
```

**Rule 6** (Variable declaration):
Finally to make our operator declarations work, we create variables, one for each sort named *aXXX*.
So, the ASN.1 module defined in Rule 5a, will have these extra lines of code when it transforms into CafeOBJ:
```
CafeOBJ:
var atypename : Typename
var asubtypename₁ : subtypename₁
var asubtypename₂ : subtypename₂
...
var asubtypenameₙ : subtypenameₙ
```

**Rule 5b** (Equation declaration):
Finally, we can now declare the equations that complement the operators we described in Rule 5a using the variables we created in Rule 6. So, the ASN.1 module defined in Rule 5a, will have the following extra lines of code when it transforms into CafeOBJ:
```
CafeOBJ:
eq typename(asubtypename₁, asubtypename₂, ...,
asubtypenameₙ) = atypename .
eq returnsubtypename₁(typename(asubtypename₁,
asubtypename₂, ..., asubtypenameₙ)) = asubtypename₁ .
eq returnsubtypename₂(typename(asubtypename₁,
asubtypename₂, ..., asubtypenameₙ)) = asubtypename₂ .
...
eq     returnsubtypenameₙ(typename(asubtypename₁,
asubtypename₂, ..., asubtypenameₙ)) = asubtypenameₙ .
```

If these operators are fed with the main sort name, they can provide as output each of the sub sorts. For example, let's assume a Client declaration, with 'Client' being the main sort name, and 'FirstName', 'LastName' and 'Address' being sub sorts of 'Client'. These observational operators are simply tools that can help us answer questions such as:

- What is the Last Name of a given Client?
- What is the Address of a given Client?

A very important part of the translation is the handling of hidden sorts (sorts that represent the set of system states). In many CafeOBJ specifications the top sort tends to represent the field space and it's usually the hidden sort.

We saw how ASN.1 keeps everything modularized and how it does not really matter in what order we write our modules. In each ASN.1 specification there is a top-level type. It's usually the one that uses everything else, directly or not. There's a clean way to declare that in ASN.1, using the "ABSTRACT-SYNTAX" notation to identify this top-level type.

```
name-abstract-syntax ABSTRACT-SYNTAX ::=
    {top-leveltypename IDENTIFIED BY ... }
```

But we can't be sure that ASN.1's top-level type will be the hidden sort for the CafeOBJ equivalent code since that depends heavily on the programmer's intentions. We can assume that there is no way for a computer program to guess what would the hidden sort be when it translates code, we would need to give it an extra input manually. So, we do feed the engine manually, with what we want to be the hidden sort.

## 4.3. Translating ASN.1 Specifications into CafeOBJ

Feeding our proposed software environment with an ASN.1 specification, gets us a basic set of CafeOBJ modules as output, that will act as a base for our verification process. We will have to ensure the validity of the translation before we continue with the verification of critical system properties; ensure that the produced set of CafeOBJ modules is an implementation of the ASN.1 specification.

We assume the following property: given an ASN.1 specification, the generated (by our set of transformation rules) CafeOBJ set of modules is an implementation of the ASN.1 specification. The verification of this property is vital to the success of our project.
To do that, we will formalize the set of transformation rules (translator) in CafeOBJ. The 'translator' formalized in CafeOBJ is called *ASN.1-trans*. Of course that means that we must also formalize ASN.1 and CafeOBJ specifications into CafeOBJ as depicted in the work of Senachak et al. [2005]

### 4.3.1. ASN.1-trans translator
Here we will describe the relation between ASN.1-Cafe and Cafe-OTS as a set of transformation rules as in Barlas et al. [2010].
Let's keep in mind that while ASN.1 has way more potential to model every sort of exchanged messages and gives access to virtually any possible data type, we shall only give emphasis to the Boolean and numeric/key data types, assuming those are important for our goal; verification of protocols.
Also, not every ASN.1 specification can be converted into a CafeOBJ/OTS set of modules automatically. It is vital that we provide some additional manual input and code rewriting to make it work, but still the translator that we describe will be doing most of the hard work.
All of the transformation rules we described earlier will be modelled here in CafeOBJ code. We have a set of transformation rules for Sequences (or the rest of the ASN.1 types), basic data types (bool, int, etc) etc, modelled as operators. Given for example a Sequence, we have `sequence2module` that creates the CafeOBJ module named as the sequence name, defines the main sort and creates the main operator declaration and equation, and finally a variable with the name aXXX (XXX stands for sequencename) `subseq2subsort` which declares sub sorts under the sequence sort name, operator declarations and equations, and finally a variable for each such sort named accordingly.

We define an operator called `translate` that takes an ASN.1 specification AnASN (denoted by a sort ASN) and generates a CafeOBJ 'program' ACafeOBJ (denoted by sort CAFEOBJ).

The verification process that we will follow is similar to the one in the work of Senachak et al. [2005] The condition that we want to verify that holds is that given an ASN.1 specification (*B*), the ASN.1-Cafe program *A* = translate(*B*) is always an implementation of *B*. To prove that *A* is an implementation of *B* we need to show that there exists a refinement relation [Lynch 1996] (called *R*) over states of A and B provided that i) for an arbitrary initial state *s* of *A*, there exists an initial state *u* of *B* such that (*s*, *u*) є *R* and ii) For an arbitrary reachable state *s* of *A* and an arbitrary method *m* of *A*, there exists a reachable state *u* of *B* and a sequence of transitions (an execution of an action *a*) of *B* such that (*s*, *u*) є *R* and (*m*(*s*), *a*(*u*)) є *R*, where a given *m* is translated from an action *a*. This part is yet to be completed.

### 4.4. Property Verification

As we explained already, we now have a basic set of CafeOBJ modules derived from the original ASN.1 specification. But in order to check the validity of critical system properties, we will have to do a few modifications to the code, so that it resembles more the OTS/Proof score methodology [Ogata et al. 2006; Futatsugi 2006; Ogata et al. 2003].

Since ASN.1 "has no knowledge of system state", we have to somehow incorporate this into our resulting specification. The state space, as declared by the hidden sort, should change any time an action (defined by action operators) takes place. We have to create operators (transitional operators) that take the current state space, apply the action, and then create the new state space as it is defined by the action modification. We also have to create operators that measure/report properties of the system state at any point (observational operators). These operators are the ones that can tell us what changes each time an action happens.

When we do that, the specification is now at OTS standards, so all we need to do then is to insert into the specification a property that has got to hold in every reachable system state (a property that is derived from the original protocol design), and for each such state we write proof passages (proof scores) that force the system to reach that state and then check for the validity of our property.

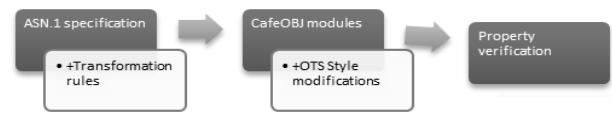

**Figure 7. Proposed transformation scheme**

Figure 7 displays the full proposed transformation scheme; assuming an ASN.1 specification, we apply our transformation rules (as seen on 4.2) to take a set of CafeOBJ modules and by making the appropriate changes to the resulting code, so it has the OTS format (3.3 and 4.4) we are ready to use the OTS/Proof score methodology in order to proceed with property checking.

### 4.5 Potential Applications

Although our proposed software environment could be used with virtually any ASN.1 specification, we believe that it would benefit more protocols that need to have their peers to be able to securely communicate with each other. Network protocols fall under this category. Since ASN.1's primary function is to model the messages that are exchanged during a point to point communication, lots of network protocol designers write ASN.1 specifications for their work. Getting them to work on CafeOBJ, makes it easy for a CafeOBJ user to further explore the security measures that are being taken on pre-coding stage, and spot flaws or verify the correctness of the protocol. For instance, knowing the format of the exchanged data over a Wi-Fi network secured with some new protocol, we can extract a CafeOBJ specification of the protocol and see if any eavesdropper can breach into the system or not.

We think of this idea as a step towards the bridging of two different worlds; the world of formal methods/formal verifications and the current industry techniques of development and release of a security protocol. Any sort of such automation makes easier for the industry world to adopt formal methods into their research & development.

### 5. CASE STUDY: A BANKING ACCOUNT STRUCTURE

In this case study, we will consider a banking account structure written in ASN.1, as seen in the work of Barlas et al. [2010]. Then we use all of the aforementioned transformation rules to produce a working set of CafeOBJ modules, and finally prove some trivial yet critical properties. We will then modify this resulting set of CafeOBJ modules so that it conforms to the OTS standard, as we previously stated, and then find out some basic, crucial properties that we want our designed system to hold at any case. This is the proposed path that a designer should follow, so that he can extend an existing ASN.1 specification allowing validation of his system We will model the data needed for a potential client that can open a bank account, the actual account, and among other minor things, the two actions (withdraw money from the account, and deposit into) and a query that returns the current balance of the x account.

### 5.1 ASN.1 Module

We will model a potential Client's information as an ASN.1 Sequence type, holding information such as name and address data, with ClientID being a unique identifier for each client. The ASN.1 module for the 'client' has been given earlier (Figure 3). Each and every account is identified by its IBAN (International Bank Account System), an account owner (a bank client - expressed by Client) and the current balance of it. To conclude the specification of Account we will also have to declare the encapsulated data type 'balance', as we haven't done that yet.

The two actions that can happen in our bank account system are: depositing money into an account, and withdrawing from it. Each one of those transactions is bound to a specific account as declared by account data type, information for the client who performed that transaction as expressed by client data type (we will need that for some verification purposes), the date on which the transaction took place and last but not least the amount of money deposited or withdrawn from the account.

All these would look like this on ASN.1:

```
Balance ::= SEQUENCE {
       iban            Iban,
       amount          REAL   }
Date ::= NumericString (SIZE (8)) -- DDMMYYYY
Deposit ::=   SEQUENCE {
       account Account
```

```
        clientid     ClientID,
        date         Date,
        amount       Real }
Withdraw ::=  SEQUENCE {
        account Account
        clientid     ClientID,
        date         Date,
        amount       Real }
```
**Figure 8. Balance & Deposit/Withdraw**

## 5.2 CafeOBJ Code
Let's see this CafeOBJ code, starting from the Client module, which is of type 'Sequence'.
Using the transformation rules we described earlier for Client structure, here's the resulting CafeOBJ code:

```
mod CLIENT{
[Client > ClientID FirstName LastName Address
PostCode City Country]
op client : ClientID FirstName LastName Address
PostCode City Country -> Client
op returnclientid : Client -> ClientID
op returnfirstname : Client -> FirstName
var aclient : Client
var aclientid : ClientID
var afirstname : FirstName
var alastname : LastName
…
eq client(aclientid, afirstname, alastname,
aaddress, apostcode, acity, acountry) = aclient .
eq returnclientid(client(aclientid, afirstname,
alastname, aaddress, apostcode, acity, acountry))
= aclientid .
eq returnfirstname(client(aclientid, afirstname,
alastname, aaddress, apostcode, acity, acountry))
= afirstname . }
```
**Figure 9. Client module**

We've skipped the rest of the operators and their equations, as their declarations are similar to the ones above. Let's proceed with the main module, the Account. Our hidden sort here will be *Account* so we will declare that in '*[ ]*' and also we will import the CafeOBJ included Integer module ("Int").
All ASN.1 data types that use the hidden sort, will be nested inside that module. So here besides *Account* we also have, *Balance*, *Deposit* and *Withdraw*, all inside the same module. This is what the automatic translation would give us.

```
mod ACCOUNT {
pr(INT + CLIENT )
*[Account]*
[Iban ClientID Balance]
[Deposit > ClientID Iban Date Int]
[Withdraw > ClientID Iban Date Int]
op withdraw : Account ClientID Date Int ->
Withdraw
op returnaccount2 : Withdraw -> Account
op returnclientid3 : Withdraw -> ClientID
op returndate2 : Withdraw -> Date
op returnamount2 : Withdraw -> Int
eq withdraw(aaccount, aclientid, adate, aint) =
awithdraw .
eq returnaccount2(withdraw (aaccount, aclientid,
adate, aint)) = aaccount .
eq returnclientid3(withdraw (aaccount, aclientid,
adate, aint)) = aclientid .
eq returndate2(withdraw (aaccount, aclientid,
adate, aint)) = adate .
eq returnamount2(withdraw (aaccount, aclientid,
adate, aint)) = aint . }
```
**Figure 10. Account module**

We now have the specification of the protocol (in this test case, a bank account system) in a language that we can extend trying to check/validate/falsify some properties of it.

## 5.3 Adding Some Features
This resulting set of CafeOBJ modules now serves as the backbone of our banking account system specification. We now need to model the changes that can happen to system states, bringing the specification closer to OTS requirements. Then we'll need to find some extra (security) properties that we'll incorporate into this system. Some basic things we could add:
We want to allow someone to withdraw money from an account only if the requested amount is not greater than the current available amount of the account. Also, one can only withdraw money from an account, if he is the owner of the account.
Those properties were of course not incorporated into the ASN.1 specification, but any programmer who deals with a protocol knows what security properties are to be maintained at all times. So, we will now modify the *Account* module as follows: We add an initial state to the system, that currently an account holds no money. We also simplify a bit the balance operator just for the sake of readability.

```
op init : -> Account
op balance : Account -> Int
eq balance(init) = 0 .
eq balance(deposit(aaccount, aclientid, adate,
aint)) = balance(aaccount) + aint .
```
**Figure 11. Adding initial state and balance equation**

The deposit action is free of any specific properties, since anyone can deposit any amount to any account. The act of depositing creates a new version of 'account' so all we have to do is to observe the new value, calculating the new balance. The operator 'balance' acts as an observational operator, where given an account A, balance returns the current balance. Operator 'deposit' (and 'withdraw') is an action operator that changes the current state. The same thing applies to the withdraw operator, but this time we also want to add a checker first, before the withdrawing can actually happen.
To model these two properties in CafeOBJ, we will have to do a few minor changes and additions as following:

```
ceq
balance(withdraw(aaccount,aclientid,adate,aint)) =
balance(aaccount) - aint
        if (balance(aaccount) >= aint and
returnclient(aaccount) == aclientid).
ceq withdraw(aaccount,aclientid,adate,aint) =
aaccount if not(balance(aaccount) >= aint) .
```
**Figure 12. Inserting a balance/owner check before withdrawal**

## 6. CONCLUSIONS
We presented a small subset of the encoding rules that can be used to convert ASN.1 specifications into CafeOBJ. Also, we sketched some ideas towards the verification of our translation. We hope in the future to create and present you a fully automated piece of software that can do this procedure.
Bringing closer two different cultures is what makes an automated translator so promising as it combines the popularity of ASN.1

and the powerful proving engine of CafeOBJ. It is a step towards the adoption of algebraic specification methods by protocol designers who are already capable of using ASN.1. Using this, a protocol can be studied, verified and validated, at the pre-coding stage, when errors are not critical, allowing more stability when the protocol finally reaches the end-user. The technique can be applied either for existed protocols or for new ones. The full correctness of the translator is expected to be proved at a later stage.